\documentclass[a4paper]{jpconf}
\usepackage{graphicx}
\begin{document}
\title{Neutrinos, Weak Interactions, and r-process Nucleosynthesis}

\author{A.B. Balantekin}

\address{University of Wisconsin, Department of Physics, Madison, 
WI  53706, USA}

\ead{baha@physics.wisc.edu}

\begin{abstract}
Two of the key issues in understanding the neutron-to-proton ratio 
in a core-collapse supernova are discussed. One of these is the behavior  
of the neutrino-nucleon cross sections as supernova energies. The other 
issue is the many-body properties of 
the neutrino gas near the core when both one- 
and two-body interaction terms are included. 
\end{abstract}

\section{Introduction}
\vskip 0.25cm
A good number of the isotopes of 
nuclei heavier than iron are produced by the rapid 
neutron-capture process (the so-called r-process) nucleosynthesis, 
the site of which is not known. R-process 
nucleosynthesis requires interaction of a large number of neutrons 
during a relatively short duration, suggesting that its sites 
are likely to be associated with explosive phenomena. 
In fact, Burbidge {\it et al} originally 
suggested that the neutrino-rich ejecta outside the collapsed core 
in a supernova could be an r-process site \cite{Burbidge:1957vc}. 
Recent observations of metal-poor stars and meteoric data suggest that 
r-process nuclei may be produced at multiple sites \cite{Qian:1997kz}. 
Possible sites for the r-process nucleosynthesis include the 
neutrino-driven wind in a core-collapse supernova 
\cite{Woosley:1994ux,witti} or binary neutron-star systems 
\cite{Rosswog:1998hy}. 

Neutrino interactions play a very crucial role both in the dynamics 
of core-collapse supernovae and in the r-process nucleosynthesis 
\cite{Balantekin:2003ip}. Isotopic yields of r-process nucleosynthesis 
are determined by the neutron-to-proton ratio, n/p \cite{Meyer:1998sn}. 
There are two key issues in understanding how this ratio changes: 
i) Neutrinos and 
antineutrinos streaming out of the core interact both with nucleons 
and the seed nuclei. Since these interactions determine the n/p ratio, it 
is crucial to understand neutrino-nucleon cross sections in considerable 
detail. ii) Before these neutrinos reach the r-process region they 
may undergo matter-enhanced 
neutrino oscillations as well as coherently scatter over other neutrinos. 
Many-body behavior of this neutrino gas is not understood, but may have 
significant impact on the r-process nucleosynthesis. Here I briefly 
discuss our recent work addressing these two issues. 

\section{Testing Neutrino-Nucleon Interactions}
\vskip 0.25cm
The covariant matrix element for the reaction 
$\overline{\nu }_{e}+p\rightarrow e^{+}+n$ is 
\begin{equation}
\frac{G_{F}\cos \theta _{C}}{\sqrt{2}}\left\{ 
\overline{u}_{n}\left[ \gamma _{\alpha }(f_{1}-g_{1}\gamma _{5}) 
+\sigma _{\alpha \beta}k^{\beta }(f_{2}+g_{2}\gamma _{5}) 
+k_{\alpha }(f_{3}+g_{3}\gamma _{5})\right] u_{p}\right\} 
\left\{ \overline{\nu }_{\nu }\gamma^{\alpha }(1 
-\gamma _{5})\nu _{e}\right\},  \end{equation}
where $G_{F}$ is the Fermi weak coupling constant and $\cos \theta_{C} 
=0.974$ is the Cabibbo angle. 
The vector $f_{1}$, the axial-vector $g_{1}$, 
the tensor $f_{2}$ (or weak magnetism), the induced tensor $g_{2}$, 
the induced scalar $f_{3}$ and the induced pseudo-scalar $g_{3}$ form factors 
are included in this matrix element. 
The conserved vector current (CVC) hypothesis states that
\begin{eqnarray}
\lim_{q^2 \rightarrow 0} f_1(q^2) ~=~ 1~; \hspace{0.75cm} 
\lim_{q^2 \rightarrow
0} f_2(q^2) ~=~ \frac{\mu_p-\mu_n}{2m_N}~; \hspace{0.75cm} f_3(q^2) ~=~ 0~,
\end{eqnarray}
where $\mu _{p}-\mu _{n}=3.706$ is the difference in the anomalous 
magnetic moments of the nucleons, and $m_{N}$ is the mass of the nucleon. 

Data for neutrino-nucleon scattering at energies relevant to  
astrophysics are very scarce. One possibility to measure the cross section for 
this process is to utilize proposed beta-beam facilities. 
Beta-beams are pure $\nu_e$ or $\overline{\nu}_e$ beams produced by 
allowing radioactive ions circulating in a storage ring to 
decay \cite{Zucchelli:2002sa}.  The low-energy option  
\cite{Volpe:2003fi,Volpe:2006in}, 
where one can vary 
the Lorentz boost factor $\gamma$ of the stored ions between 7 and 14, 
is especially 
suitable to measure the energy dependence 
of the cross section in the energy range up to 100 MeV.  
The potential of low-energy beta beams for neutrino-nucleus 
scattering \cite{Serreau:2004kx,McLaughlin:2004va,Volpe:2005iy}, 
for electroweak tests of the 
Standard Model \cite{McLaughlin:2003yg,Balantekin:2005md}, and for  
core-collapse supernova physics \cite{Volpe:2003fi,Jachowicz:2006xx} 
has been previously discussed.

We recently investigated the possibility of measuring antineutrino 
scattering off protons in a water Cerenkov detector using low-energy 
beta beams \cite{Balantekin:2006ga}. We analyzed the sensitivity using 
both the total number of events and the angular distribution of the 
positrons emitted. Figure 1 shows the expected number of events at the 
detector over a period of one year as a function of the Lorentz boost 
factor with (solid-line) and without (dashed-line) the weak magnetism 
term. One observes that at the highest $\gamma$ value the weak magnetism 
term suppresses the total number of events by as much as 17 \%. We found 
that the weak magnetism form factor may be determined with better than 
several percent accuracy using the angular distribution information of 
the positrons emitted. In addition to providing a test
of the CVC hypothesis, measurement of weak magnetism terms in 
neutrino-nucleon interactions are of direct interest to astrophysics as 
these terms may play an important role in the dynamic of core-collapse 
supernovae \cite{Horowitz:2001xf}.

\begin{figure}[t]
\hskip 4.3cm
\includegraphics[width=7cm]{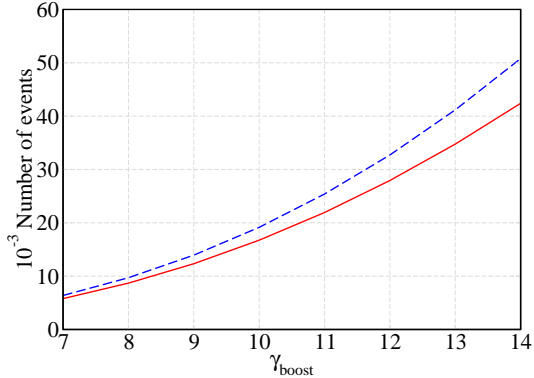}
\caption{Expected number of the $\overline{\protect\nu}_e + p 
\rightarrow e^+ + n$ events in a water Cerenkov detector over one year 
with (solid-line) and without (dashed-line) the weak magnetism term 
(From Ref. \cite{Balantekin:2006ga}).}
\end{figure}

\section{Many-Body Properties of the Dense Neutrino Gas in Supernovae}
\vskip 0.25 cm
Neutrino oscillations in a core-collapse
supernova differ from the matter-enhanced neutrino oscillations in
the Sun as in the former there are additional effects coming from both
neutrino-neutrino scattering
\cite{Pantaleone:1992eq,Qian:1994wh} and
antineutrino flavor transformations \cite{Qian:1995ua}. 
The standard MSW term describing the interaction of neutrinos with the 
background electrons is a one-body term for neutrinos. 
Exact solutions of the full neutrino many-body problem with the one-body 
MSW term and the two-body neutrino-neutrino forward scattering term have not 
yet been found. 

For two neutrino flavors, i.e. the
electron neutrino, $\nu_e$, and a combination of muon and tau
neutrinos, called $\nu_x$, the one-body term in the neutrino many-body 
Hamiltonian is \cite{Balantekin:2006tg}
\begin{equation} 
\label{a1}
H_{\nu} = \int d^3p \frac{\delta m^2}{2p} \left[ \cos{2\theta}
J_0(p) + \frac{1}{2} \sin{2\theta} \left(J_+(p)+J_-(p)\right)
\right] - \sqrt{2} G_F  \int d^3p N_e J_0(p) , 
\end{equation}
where $N_e=n_{e^-}-n_{e^+}$ is the net electron density;
$\theta$ is the vacuum mixing angle, and
$\delta m^2 = m_2^2 - m_1^2$. In writing Eq. (\ref{a1}) we introduced 
the operators 
\begin{equation}\label{a2}
J_+(p)= a_x^\dagger(p) a_e(p), \ \ \ \
J_-(p)=a_e^\dagger(p) a_x(p), \ \ \ \
J_0(p)=\frac{1}{2}\left(a_x^\dagger(p)a_x(p)-a_e^\dagger(p)a_e(p)
\right) ,
\end{equation}
where we used the creation and annihilation operators for $\nu_e$ 
and $\nu_x$. The operators in Eq. (\ref{a2}) satisfy the commutation 
relations
\begin{equation}
\label{a3}
[J_+(p),J_-(q)] = 2 \delta^3(p-q)J_0(p), \ \ \
[J_0(p),J_\pm(q)] = \pm \delta^3(p-q)J_\pm(p).
\end{equation}
These operators describe $N$ mutually commuting SU(2) algebras, where 
$N$ is the number of allowed values of neutrino momenta. In fact, 
the evolution operator for the standard MSW problem of neutrinos,
mixing with each other and interacting with background electrons, can be
written down exactly using this algebraic ansatz. In addition, there is 
a two-body term in the 
Hamiltonian describing neutrino-neutrino forward scattering: 
\begin{equation}
H_{\nu \nu} = \frac{\sqrt{2} G_F}{V} \int d^3p \> d^3q \> 
(1-\cos\vartheta_{pq}) 
\> {\bf J}(p) \cdot {\bf J}(q) .
\end{equation}
where $V$ is the quantization volume and $\vartheta_{pq}$ is the angle 
between neutrino three-momenta p and q. 

One can calculate the path integral for the many-body problem using the SU(2) 
coherent states
\begin{equation}
|z(t)\rangle = \exp{\left(\int d^3p
z(p,t) J_+(p) \right)} \prod_{p}
a_e^\dagger(p)|0 \rangle .
\end{equation}
Path integral representation of the matrix element of the
evolution operator calculated with $H_{\nu} + H_{\nu \nu}$ is given by
\begin{equation}
\label{a4} \langle z'(t_f)|U|z(t_i) \rangle = \int D[z,z^*] \,
e^{iS[z,z^*]}, 
\end{equation}
where the action functional is 
\begin{equation}
\label{a5} S[z,z^*]= \int_{t_i}^{t_f}dt \langle z'(t_f) | 
i\frac{\partial}{\partial t}-H_\nu-H_{\nu\nu} | z(t_i) 
\rangle -i \log
\langle z'(t_f)|z(t_f) \rangle.
\end{equation}

The path integral given in Eq. (\ref{a4}), to date, has not been evaluated 
exactly. However, it is possible to evaluate it using the stationary path 
approximation and calculate the stationary path $z_{\rm sp}(p,t)$ that 
minimizes the action functional \cite{Balantekin:2006tg}. By interpreting 
this stationary path as the ratio of the one-body neutrino wavefunctions, 
i.e. 
\begin{equation}
\label{a6}
z_{\rm sp}(p,t)=\frac{\psi_x(p,t)}{\psi_e(p,t)}
\end{equation} 
and imposing the normalization condition
\begin{equation}
\label{a7} |\psi_e|^2+|\psi_x|^2=1 ,
\end{equation}
one can relate results obtained using stationary path approximation to 
the commonly utilized Schrodinger-type equation such as the one used in Refs. 
\cite{Fuller:2005ae,Friedland:2003dv,Duan:2005cp}. 
Numerical solutions of this approximate 
Schrodinger-type nonlinear equation are still 
not easy to obtain as discussed in Refs. 
\cite{Pastor:2002we,Balantekin:2004ug,Friedland:2006ke,Duan:2006an}. 

A description of neutrino physics with two flavors is exact only when 
the third mixing angle, $\theta_{13}$, is zero \cite{Balantekin:1999dx}. 
However, it is possible to incorporate both the antineutrinos (which are 
crucial to nucleosynthesis in core-collapse supernovae) and the presence 
of three flavors in this algebraic formalism \cite{Balantekin:2006tg}.

\section{Conclusions}
\vskip 0.25cm 
It should be emphasized that neutrino oscillations significantly 
impact r-process 
nucleosynthesis only if different flavors initially have well-pronounced 
energy differences. The precise energy hierarchy of the neutrinos 
depends on the microphysics of their production 
\cite{Raffelt:2001kv,Keil:2002in}. 
This microphysics is dominated by the inelastic 
neutrino-nucleon interactions.  

In a core-collapse supernova, as the alpha particle 
mass fraction increases, free nucleons
get bound in alphas and, because of the large binding energy of
the alpha particle, cease interacting with neutrinos. (This is called 
``alpha effect'' \cite{Fuller:1995ih}). Electron neutrinos radiated from 
the proto-neutron stars are typically 
too energetic to prevent the alpha effect (and the following demise 
of r-process nucleosynthesis) 
in most cases. One way to get around this issue is to convert active 
electron neutrinos into sterile ones 
\cite{McLaughlin:1999pd,Caldwell:1999zk,Fetter:2002xx}. Inclusion of 
sterile neutrinos in the algebraic approach described above will be 
discussed elsewhere. 

Finally, it should be pointed out that the neutrino gas is not necessarily 
present in all cases. If, instead of a proto-neutron star, a black-hole 
is formed, then the neutrino flux emission 
may be truncated \cite{Beacom:2000qy}.  Note that 
it may be possible to find signatures of such a neutrino-flux truncation 
in the fossil record of the isotopes produced in the r-process 
nucleosynthesis \cite{Sasaqui:2005rh}. 

\ack
\vskip 0.25cm
This work was supported in part by the U.S. National Science
Foundation Grant No. PHY-0555231 and in part by the University of
Wisconsin Research Committee with funds granted by the Wisconsin
Alumni Research Foundation.

\end{document}